\documentstyle[11pt,newpasp,twoside,epsf]{article}

\begin{document}

\title{VIMOS IFU as a tool to observe the cores of high z clusters}
\author{D. Rizzo, on behalf of the VIRMOS Consortium}
\affil{Brera Astronomical Observatory, Via Bianchi 46, I-23807 Merate (LC), Italy}

\begin{abstract}
Research on the core of medium and high z clusters of galaxies can derive great benefits from integral field spectroscopy, and a key role in this respect will be played by the Integral Field Unit (IFU) being developed by the VIRMOS Consortium as part of the VIMOS spectrograph. 
After a brief technical outline of the instrument, some of the problems to be addressed by data reduction techniques are described, and the scientific issues to which VIMOS IFU is likely to give its major contribution are pointed out.
\end{abstract}

\section{Technical overview}

The VIMOS Integral Field Unit (hereafter IFU) allows to obtain one spectrum for each resolution element of its field of view, thanks to a square array of microlenses, each one coupled to an optical fiber. Technical details about VIMOS IFU can be found in Prieto et al. (2000). See also Le F\`evre et al. (1998).

VIMOS IFU can be used in various configurations, outlined in Table \ref{tab:ifuconf}

\begin{table}[htbp] \label{tab:ifuconf}
\caption{IFU configurations}
\begin{tabular}{cccc}
\tableline
Field of view & Spatial resolution & Spectral resolution & Number of fibers \\
\tableline
54\arcsec $\times$ 54\arcsec & 0\farcs67 & 250 & $80 \times 80$ \\
27\arcsec $\times$ 27\arcsec & 0\farcs33 & 250 & $80 \times 80$ \\
27\arcsec $\times$ 27\arcsec & 0\farcs67 & 700, 2500 & $40 \times 40$ \\
13\arcsec $\times$ 13\arcsec & 0\farcs33 & 700, 2500 & $40 \times 40$ \\
\tableline
\tableline
\end{tabular}
\end{table}

\section{Reducing IFU data}

IFU data reduction presents new challenges and requires to set up new data reduction techniques with respect to ``traditional'' MOS data.

Two main aspects that have to be taken into account for the data reduction of every IFU are the high density of spectra on the detector, causing some degree of crosstalk (overlapping of light from neighbouring spectra) and the sky subtraction, demanding special care, since the sky must be evaluated using different fibers for different observations. For VIMOS IFU in particular a third aspect is the high number of spectra collected in a single exposure (up to 6400 spectra on four 2K$\times$4K CCDs), requiring at least a partial automatization of the reduction process.

Once data have been reduced, exploiting their scientific content is even more challenging. For this reason a European network has been set up, formed by all the European institutes involved in IFU instrument development, with the purpose to provide the European astronomical community with powerful 3D data analysis tools.

\section{Doing science with IFU}

The fact that an IFU can obtain spectra for all the objects in the field of view at once makes it much more efficient than a MOS in very crowded field, where complete sampling is required. This is especially true in the case of high-z clusters, where the numbers of interlopers is roughly 90\% of all the field objects.

This can have great advantages regarding spectroscopic confirmation of clusters and dynamical studies of cluster cores, where just one IFU observation (or a few ones, if using the smaller size and higher resolution configuration) could obtain the same result as several MOS exposures. It is clear that medium and high-z clusters, being of a smaller angular size, are ideal targets for IFU observations, as shown by Figure 1.

\begin{figure}[htbp] \label{fig:ifusim}
\plottwo{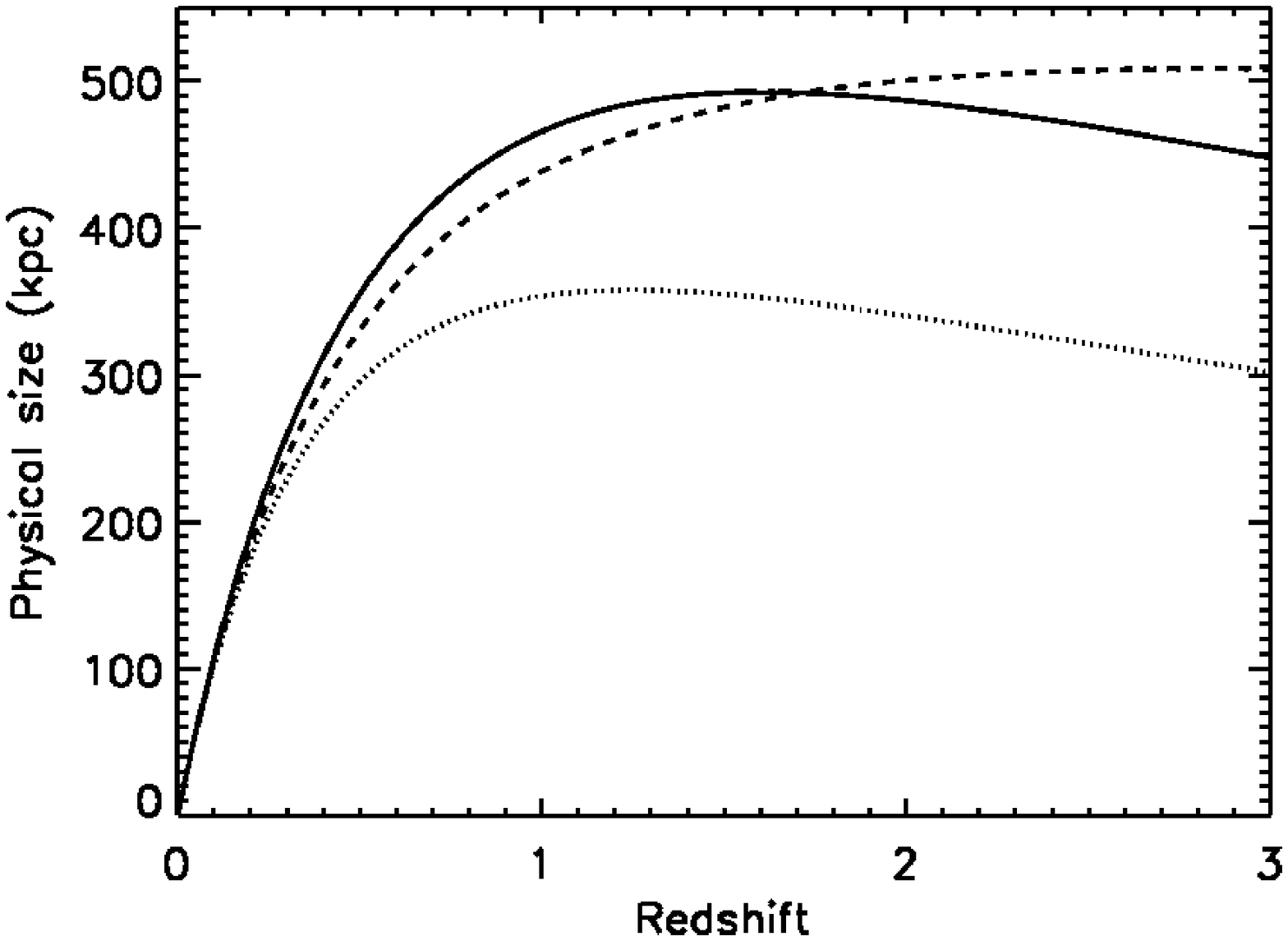}{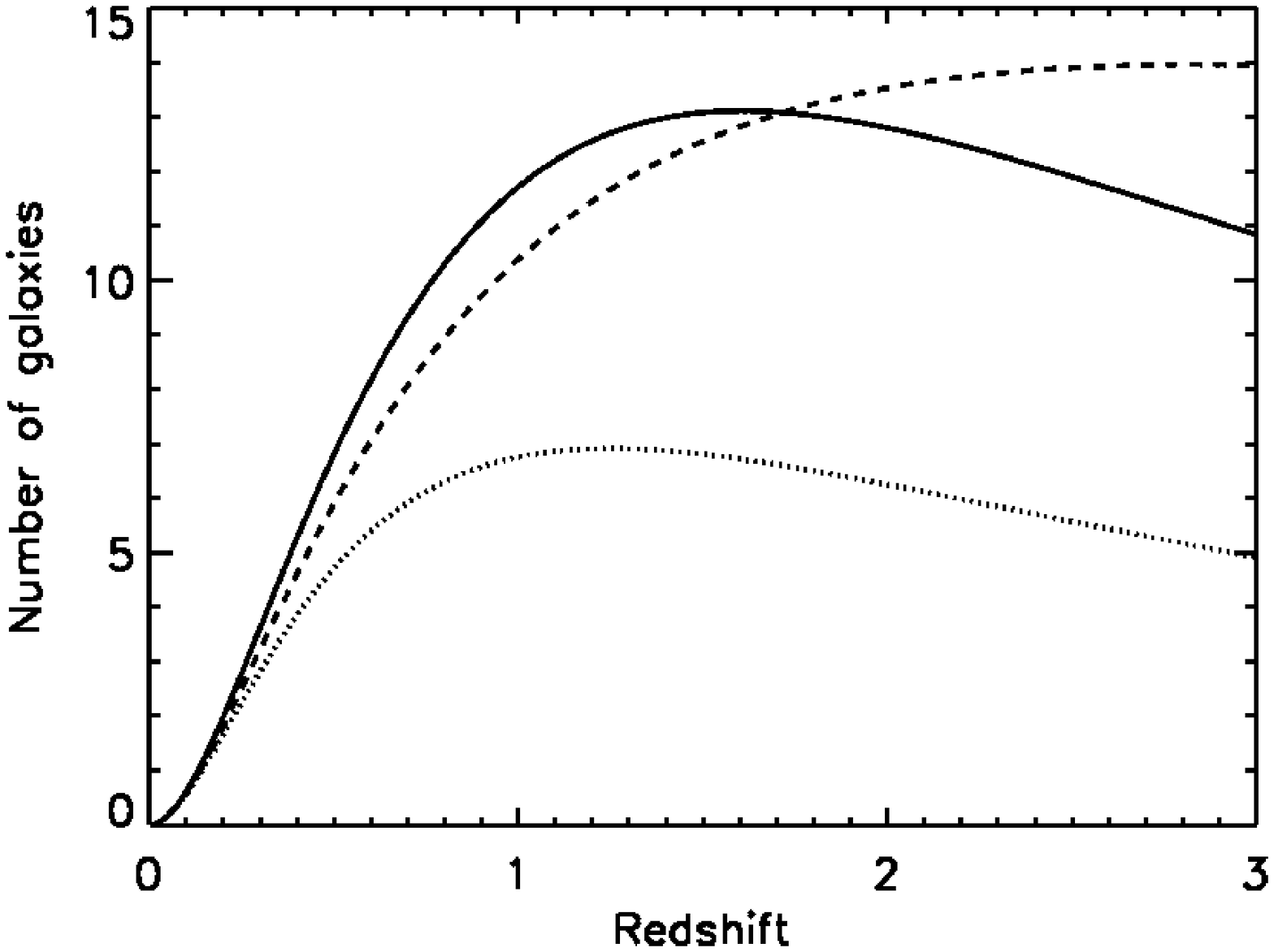}
\caption{Left: physical size of VIMOS IFU field (54\arcsec$\times$54\arcsec). Right: estimate of the number of galaxies in the field, with $m_3 < m < m_3 + 2$, belonging to a Coma-like cluster moved at different redshifts (computed using data from Biviano et al. 1995). Solid line: $\Omega_{\mathrm{M}} = 0.3, \Omega_\Lambda = 0.7$; dashed line: $\Omega_{\mathrm{M}} = 0.1, \Omega_\Lambda = 0$; dotted line:  $\Omega_{\mathrm{M}} = 1.0, \Omega_\Lambda = 0$.}
\end{figure}

Systematic IFU observations of cluster cores will be part of the VIMOS-VLT Deep Survey (Le F\`evre et al. 2000).

\end{document}